\documentclass{sig-alternate-2013}

\usepackage{graphicx}
\usepackage{xcolor}
\graphicspath{ {./images/} }

\permission{Copyright is held by the author/owner(s).}
\conferenceinfo{WWW'16 Companion,}{April 11--15, 2016, Montr\'eal, Qu\'ebec, Canada.} 
\copyrightetc{ACM \the\acmcopyr}
\crdata{978-1-4503-4144-8/16/04. \\
DOI: http://dx.doi.org/10.1145/2872518.2890534}

\begin{document}

\newcommand{\xhdr}[1]{\vspace{2mm}\noindent{{\bf #1}}}

\title{Fashionista: A Fashion-aware Graphical System \\for Exploring Visually Similar Items}

\numberofauthors{1} 
\author{
\alignauthor
Ruining He\thanks{Both authors contributed equally to this work.}, Chunbin Lin\footnotemark[1], Julian McAuley\\
       \affaddr{University of California, San Diego}\\
       \affaddr{La Jolla, California, U.S.A.}\\
       \email{\{r4he, chunbinlin, jmcauley\}@cs.ucsd.edu}
}

\maketitle
\begin{abstract}
To build
a fashion recommendation system,
we need to help users retrieve
\emph{fashionable items} that are visually similar to a particular query, for reasons ranging from searching alternatives (i.e., substitutes), to generating stylish outfits that are visually consistent, among other applications. In domains like clothing and accessories, such considerations are particularly paramount as the visual appearance of items is a critical feature that guides users' decisions.
However, existing systems like \emph{Amazon} and \emph{eBay} still rely mainly on keyword search and recommending loosely consistent items (e.g.~based on co-purchasing or browsing data), without an interface that makes use of \emph{visual information} to serve the above needs.

In this paper, we attempt to fill this gap by designing and implementing an image-based query system, called \emph{Fashionista}, which
provides a graphical interface to help users efficiently explore those items that are not only visually similar to a given query, but which are also \emph{fashionable}, as determined by visually-aware recommendation approaches.
Methodologically, \emph{Fashionista} learns a low-dimensional visual space as well as the evolution of fashion trends from large corpora of binary feedback data such as purchase histories of \emph{Women's Clothing $\&$ Accessories} from \emph{Amazon}, which we use for this demonstration.

\end{abstract}


\section{Introduction}
With the proliferation of e-commerce and online recommendation services, users increasingly rely
on intelligent systems to explore large corpora of items, in order to discover products to purchase, movies to watch, news to read, etc. When it comes to domains like clothing recommendation---where visual factors play a key role---users' decisions depend largely on the visual attractiveness (or `fashionability') of the products in question. Therefore we argue that an interactive, 
visually-aware
interface for exploring items can be of significant value in such domains.

Visual similarity entails specific notions of `visual compatibility' between items, such as substitutable and complementary products \cite{VisualSIGIR}. Substitutes are similar items that are interchangeable with one another, e.g.~one pair of shoes for another. In contrast, complements are items that can visually `go together,' such as a t-shirt and a matching pair of jeans. We hope to help users retrieve and explore items that are potential substitutes or complements for items they are currently considering.
Retrieving visual substitutes can help users to discover items that may be preferable in terms of their price, popularity, brand, or even items which are considered more fashionable by the community. Alternately, retrieving visually complementary items can help users to find items with consistent styles, and to generate sets of items (i.e., `outfits').

The major challenges for a system to provide such support lie in the ability to (1) use complex similarity measures that are capable of capturing human notions of visual similarity (i.e.,~what humans consider to be substitutable and complementary), and (2) index, query, and retrieve visually similar items from a large corpus (e.g.~millions of items) efficiently.

Furthermore, this problem is challenging as fashions are continually evolving over time, such that temporal information is key to determining what is currently considered `fashionable.'
This means that in addition to modeling large corpora of users' interactions with items, we must also model their temporal dynamics in order to capture the notion of `fashion trends.'
Our main goal is to address these challenges and design a system (which we title \emph{Fashionista}) to provide a user-friendly graphical interface to help users find \emph{fashionable} items that are visually similar to a given query.

In contrast to current approaches,
existing systems like \emph{Amazon's} own recommendation interface\footnote{http://www.amazon.com/} don't help users to retrieve visually consistent items, though they do recommend items that have been frequently bought/viewed together, which may be correlated with visual compatibility. 
Alternately, existing image search engines (like \emph{Google Images}\footnote{https://images.google.com/}) are based either on keyword search or traditional image similarity comparison; however they are not optimized specifically to capture human notions of `visual style,' i.e., they are not trained on users' interactions with items as ours is. To the best of our knowledge, none of these existing systems provides an interface that can help users evaluate the fashionability of items as \emph{Fashionista} does.

Methodologically, \emph{Fashionista} is based on state-of-the-art visual models that are able to learn the temporal evolution of fashion trends from large corpora of binary feedback data with images and timestamps \cite{VisualSIGIR,HeMcA16a,HeMcA16}. For visualization and efficient querying, all items in the corpus are embedded into a low-dimensional visual space (`style space') such that nearby items (i.e., neighbors) have been 
evaluated 
similarly (in terms of appearance) by the users in the corpus. More specifically, our contributions are summarized as follows:
\begin{itemize}
\item We propose a demonstration of a system, \emph{Fashionista}, which provides a novel graphical interface for exploring visually similar items within an automatically learned visual space. 
\item \emph{Fashionista} learns and visualizes the fashionability evolution of each item during the lifetime of the corpus, which can help users to find currently `stylish' items.
\item \emph{Fashionista} operates on large, real-world corpora of items. For demonstration, we use the \emph{Women's Clothing $\&$ Accessories} dataset from \emph{Amazon} \cite{VisualSIGIR}, which consists of over 0.6 million items, 1.8 million users, and 3.2 million user-item interactions spanning 11 years.
\end{itemize}

This rest of the paper is organized as follows. We first introduce \emph{Fashionista}'s architecture and interface in Section 2, followed by demonstrations of multiple 
use-cases
in Section 3. We discuss related works in Section 4 and conclude in Section 5. 

\section{Overview of Fashionista}
In this section we give an overview of the architecture
of \emph{Fashionista}, after which we discuss the unique features of our user interface.
\subsection{Architecture}
\begin{figure}
\includegraphics[width=\columnwidth]{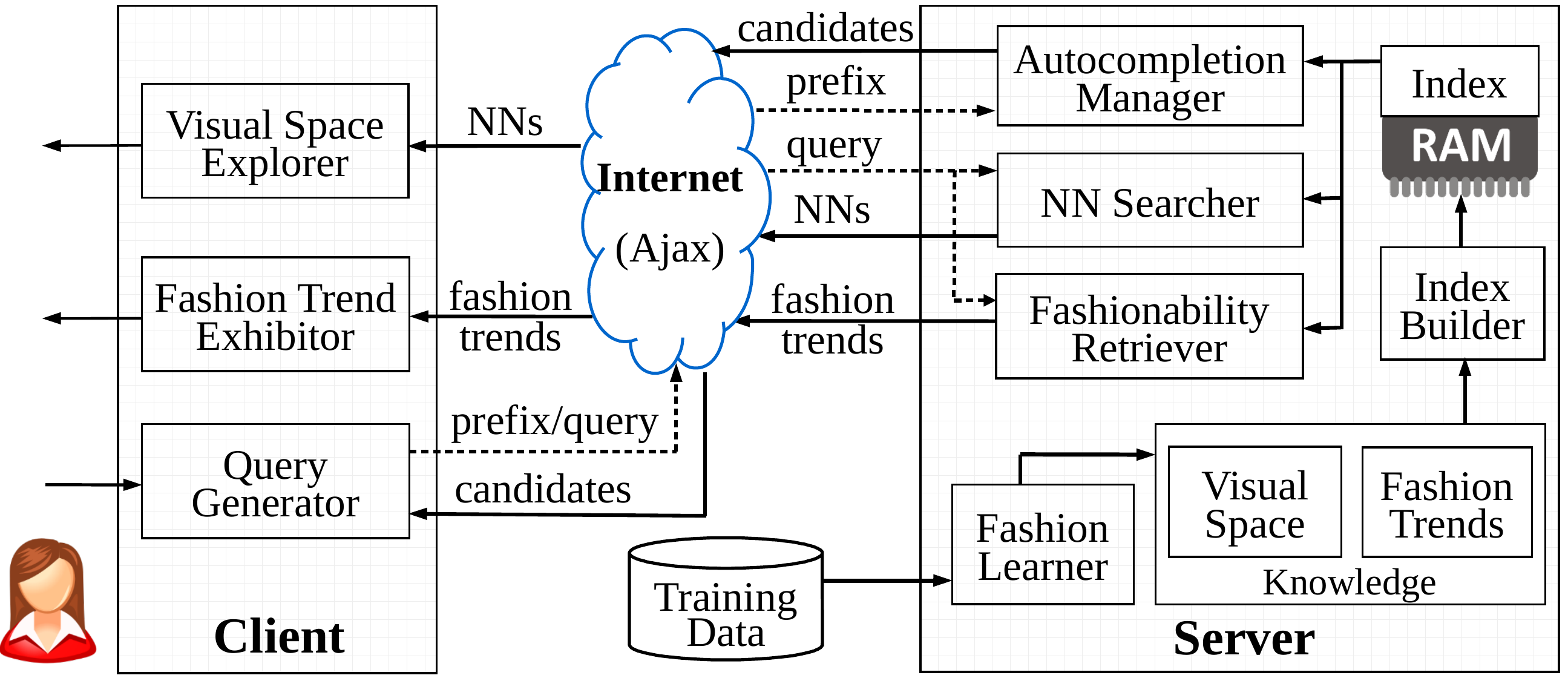}
\caption{Architecture of \emph{Fashionista}. }
\label{fig:architecture}
\end{figure}

The system architecture of \emph{Fashionista} is presented in Figure \ref{fig:architecture}. On the server side, an \textit{Index Builder} creates in-memory indices on the system's `knowledge', i.e., the visual space and fashion trends produced by the \textit{Fashion Learner}. On the client side, a query is created by the \textit{Query Generator} with the help of the \textit{Autocompletion Manager}, which provides a list of candidate queries. Once a complete query is submitted from the client, the \textit{Fashionability Retriever} and the \textit{Nearest Neighbor (NN) Searcher} are triggered to obtain the query item's fashionability values and its k-NNs. In particular, to obtain k-NNs, \emph{Fashionista} first looks up the category inverted index to retrieve a list of candidate items. Afterwards, \emph{Fashionista} builds a heap of items whose fashionability scores are greater than $\alpha$ and the top-k items are popped off the heap. Those results are then sent to the \textit{Fashionability Visualizer} and the \textit{Visual Space Explorer}
for visualization.

\begin{figure*}
\begin{center}
\includegraphics[width=\textwidth]{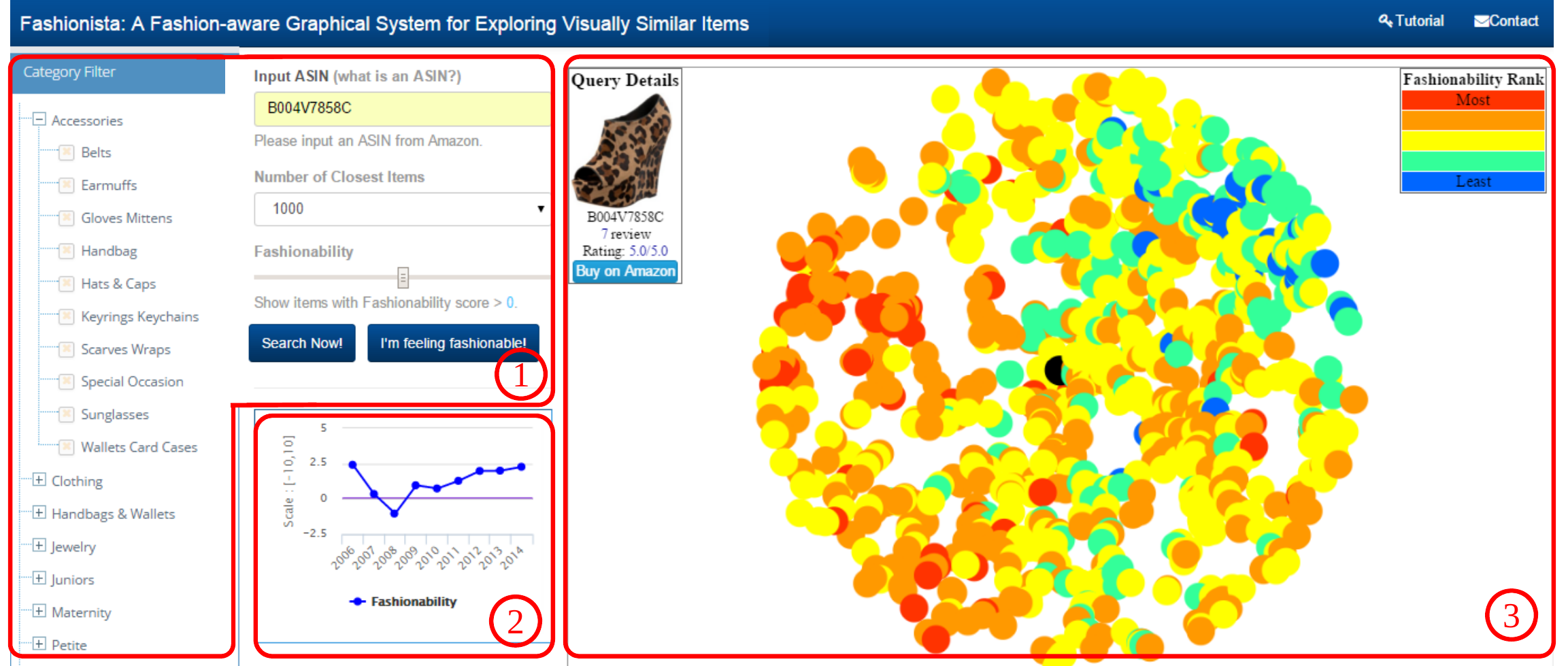}
\end{center}
\caption{A screenshot of \emph{Fashionista}. \textcircled{1} is the query generator with four components: I.~category filter (supporting multiple selections); II.~item id input text box; III.~drop-down list accepting a number of NNs to return; and IV.~
a slider to filter out unfashionable items. \textcircled{2} is the 
exhibitor showing the fashion trend of the query item. \textcircled{3} is the visual space explorer with zoom-in/out, drag and rotation features. Demo link: http://132.239.95.211:8080/demowww/index.jsp.}
\label{fig:system}
\end{figure*}

\subsection{User Interface}
The interface for \emph{Fashionista} is shown in Figure~\ref{fig:system}, which consists of three parts: the query generator, fashion trend exhibitor, and visual space explorer (corresponding to \textcircled{1}, \textcircled{2}, and \textcircled{3} in Figure~\ref{fig:system}, respectively).

\subsubsection{Query Generator}
A query generated by the \emph{Fashionista} query generator consists of four parts: (1) a list of categories chosen using the category filter component; (2) an item identifier input by the user, which can either be typed directly or chosen from a list of candidate queries provided by \emph{Fashionista}'s autocompletion manager (or selected through navigation); (3) a number $k$, which is selected from a drop-down list; and (4) a threshold value $\alpha$, which is set by sliding the window in the slider. 

Query autocompletion~\cite{lin2012lotusx} helps users to identify candidate queries when only prefixes are input. It significantly improves the efficiency of query generation since users are only required to input a few characters. In \emph{Fashionista}, the server maintains a \textit{trie} for all item identifiers in its main memory. Every time a new character is entered, an Ajax request is sent to the server which looks up the trie and returns a list of candidate items with a matching prefix. In addition, the server also returns the corresponding icon images for the candidate items to provide a preview in the item id input text box. 

\subsubsection{Fashion Trend Exhibitor}
The fashion trend exhibitor surfaces a plot of the fashionability scores of the query image from the year of 2006 to 2014, based on the model from \cite{HeMcA16a}. Technically, \emph{Fashionista} maintains an inverted index for each item to index its fashionability scores. For example, the fashionability trend in Figure~\ref{fig:system} tells the user that the shoe (with item id \textit{B004V7858C}) is gaining fashionability in recent years.

\subsubsection{Visual Space Explorer}
The visual space explorer visualizes the k-NNs of the query image within the visual space, after filtering out items not from the selected categories or with fashionability scores lower than $\alpha$. 
The k-NNs are displayed as a heat map where the color of each point indicates the fashionability of the item,\footnote{Warmer color means greater fashionability score.} as shown in \textcircled{3} in Figure~\ref{fig:system}. When the spot is clicked, detailed information like brand, price, rating, etc.~of the clicked item is displayed on a product description panel.

\begin{figure}
\begin{center}
\fbox{
\includegraphics[width=0.94\linewidth]{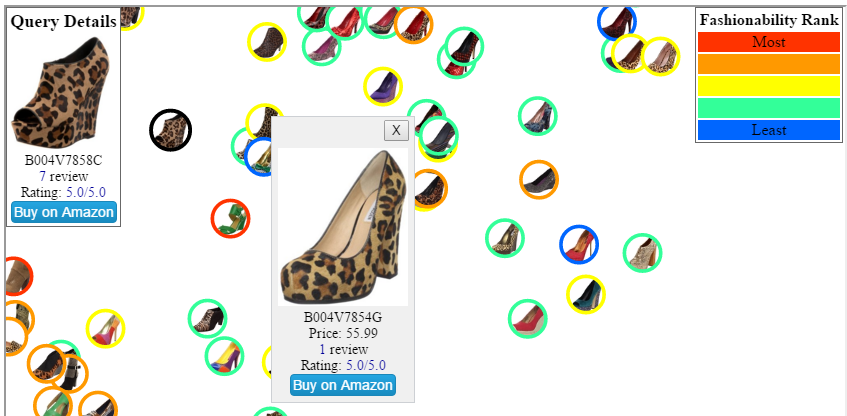}
}
\end{center}
\caption{A screenshot of the zoom-in effect of the visual space explorer. Note that (1) the detailed information of each item can be obtained by clicking its image; and (2) the color of the circled border of each image indicates the item's fashionability rank. The ranking color bar can be seen at the top-right corner.}
\label{fig:zoom}
\end{figure}

Users can explore this visual space by (1) zooming in/out and dragging the visual space explorer, and (2) clicking the 
ID
of an item on its product description panel to issue a new query. 
Figure~\ref{fig:zoom} shows an example where we zoom in the area around the query item \textit{B004V7858C}. As we can see from this figure, the nearest neighbors retrieved by \emph{Fashionista} share similar visual style with the query. Assuming the user is attracted 
to
this neighbor, they can query this item simply by clicking its ID.

\subsection{Fashion Learner}
\emph{Fashionista} is built on top of state-of-the-art visual models that are capable of learning both the dimensions of people's visual preferences as well as their temporal dynamics, i.e., fashion trends.

\xhdr{Training data.}
To capture the temporal dynamics of the general tastes of the public, our model directly learns from large datasets from \emph{Amazon}, introduced by \cite{VisualSIGIR}. This dataset temporally spans almost one decade. Training data consists of triplets of the form $(user, item, timestamp)$ denoting one purchase entry. Additionally, one image for each item is used to extract high-level visual features from a pretrained Deep Convolutional Neural Network (Deep CNN) \cite{DeepCNNArchitecture}, fed as input to the model. This is a One-Class Collaborative Filtering setting \cite{OCCF} where our model estimates users' \emph{fashion-aware} personalized ranking functions based on their past feedback.

\vfill

\xhdr{Model Specifics.} 
Our model produces a \emph{visually-aware} preference predictor that extends standard Matrix Factorization by modeling \emph{visual} dimensions and non-visual (i.e., latent) dimensions simultaneously. Here the basic idea is to uncover low-dimensional decision factors to explain user's historical purchase activities. To account for fashion dynamics, our model further captures the following phenomena: (1) items gradually gain/lose `attractiveness' in different visual dimensions over time; (2) 
users weigh visual dimensions differently as time goes by; and (3) each user's visual preferences also evolve with time. Our model uses `global' structures to capture the first two phenomena which describe visual dynamics at the community level (i.e., fashion evolution) as well as `local' structures to account for personal evolution. See our papers for further details \cite{HeMcA16a, HeMcA16}. Ultimately, our model uncovers visual dimensions and fashion dynamics by predicting users' behaviors as accurately as possible. 

Finally, the output of the model includes (1) a projection of items into an uncovered `visual space' where items
with similar styles---in terms of the way their appearance is evaluated---are 
embedded to nearby positions, and (2) fashionability evolution of each item during the time span of the dataset, among other features.
To visualize the visual space (10 dimensional in our case) on the user interface, \emph{Fashionista} further embeds it in two dimensions with t-SNE \cite{tsne}.

\section{Demonstration \& Use Cases}
In this section, we demonstrate different use-cases and real-world scenarios where our system might be used to help a user to explore the visual space. Each of these use-cases can be naturally handled by \emph{Fashionista}, and to the best of our knowledge can not be handled by existing systems.
\subsection{Querying for fashionability advice}
Rachel is hesitant to purchase a dress on \emph{Amazon}. She personally likes it but is not sure whether it is consistent with the current `fashion zeitgeist,' i.e., its fashionability. She
searches for the dress on \emph{Fashionista}. \emph{Fashionista} demonstrates the fashionability evolution of the dress's appearance in the past decade, and Rachel finds that it's gaining popularity in recent years. Armed with this information, she decides to purchase the dress immediately. 

\subsection{Searching substitutes with similar styles}
Lucy finds a pair of shoes (e.g.~\textit{B004V7858C}) on \emph{Amazon}, but wants to compare them against alternatives that are similar in appearance, but potentially have a preferable brand, price, or rating (etc.). Among items that are `frequently bought together' (as recommended by \emph{Amazon}), Lucy finds they tend to vary too much in appearance, or otherwise are not visually attractive. Using \emph{Fashionista}, she quickly retrieves hundreds of visually similar items
and finds a shoe of the same style but with a preferable brand.

\subsection{Finding complements for outfit generation}
Angelina is herself a fashionista who just bought a beautiful t-shirt on \emph{Amazon}, and she wants to find some pants and shoes that can go together with it, that is, to generate an outfit with a consistent visual style. She uses \emph{Fashionista}'s category filter to limit the search results to be within the two categories under consideration. According to the query conditions, the system retrieves the nearest neighbors of the t-shirt, according to the learned visual `style space,' each represented with different colors denoting their visual popularity. Angelina explores these items and finds several with high fashionability scores that match the style of the t-shirt.

\subsection{I'm feeling fashionable!}
Unlike the above scenarios, Christina doesn't know what she really wants, but just intends to explore some fashionable items. She clicks the `I'm feeling fashionable!' button on \emph{Fashionista}, and the system returns a randomly selected hotspot in the visual space that currently has a high popularity (fashionability). She decides to focus on fashionable coats, using the provided category filter.
Some attractive coats are surfaced, and Christina decides to buy one of them using the redirection links. 

\subsection{Querying for statistical fashion trends}
Jennifer is a fashion designer who cares about the trends of contemporary fashion. Using \emph{Fashionista}, she quickly checks the distribution of popular appearances over hundreds of thousands of clothing and accessory items. She zooms into some areas that interest her the most and observes the corresponding fashion trend evolution during the past decade.
She identifies certain trends that are likely to gain popularity in the near future and decides to design products that fit them.

\section{Related Work}
Most relevant to our work are existing e-commerce systems like \emph{Amazon}, \emph{eBay},\footnote{http://www.ebay.com/} \emph{Walmart},\footnote{http://www.walmart.com/} etc.~In spite of the wide successes achieved by such systems, they are not designed for exploration of visually relevant items. Although they are able to recommend items that are frequently bought or viewed together, they are unable to tease-apart the underlying reasons including functionality, complementarity,
visual compatibility,
and so forth. This limitation makes it difficult for these systems to address the task of recommending visually consistent items, nor do they provide an interface for efficient exploration within a visual space as we do in this work. 

There are also image search engines like \emph{Google Images} that are either based on keywords or image similarity comparison. Such systems differ from ours in two key ways. First, such systems are general-purpose and thus are not suitable for the specific applications considered here.
Second, our model maps items into a low-dimensional visual space such that nearby items have been 
evaluated 
similarly (by users) in terms of their appearance.
This improves upon traditional image similarity comparison methods by performing comparisons only on those dimensions that are discovered to be relevant to people's decisions.

\section{Conclusion and Future Work}
In this paper, we built a demo system, \emph{Fashionista}, with a user-friendly graphical interface to search and explore visually similar and fashionable items. Experimentally, we found that \emph{Fashionista} can index and search large-scale real-world corpora efficiently and effectively. In the future, we hope to enable our system to automatically generate outfits according to \emph{personalized} visual preferences and fashion trends. This will be feasible once we are able to observe users' interactions with \emph{Fashionista}.

\bibliographystyle{ieeetr}
\bibliography{sigproc} 

\end{document}